# Canonized then Minimized RMSD for Three-Dimensional Structures

*Jie Li, Qian Chen, Jingwei Weng, Jianming Wu*, Xin Xu*

Department of Chemistry, Fudan University, Shanghai, 200438, China

ABSTRACT: Existing molecular canonization algorithms typically operate on one-dimensional (1D) string representations or two-dimensional (2D) connectivity graphs of a molecule and are not able to differentiate equivalent atoms based on three-dimensional (3D) structures. The stereochemical tags on each atom are in fact determined according to established Cahn-Ingold-Prelog (CIP) rules for comparing grades, which can help to further differentiate atoms with similar environment. Therefore, a stereochemical-rule-based canonization algorithm that is capable of assigning canonical indices using 3D structural information is of great value. On top of the Schneider-Sayle-Landrum (SSL) partition-based canonization algorithm, we propose an enhanced canonization algorithm to expand its applicability. The initial index assignment rules are redesigned, so that the obtained canonical indices are compatible with the most of the common CIP Sequence Rules, which greatly eases the stereochemical assignment. Furthermore, a branching tiebreaking step is added to secure an accurate evaluation of the structural difference through the minimized root-mean-square deviation (RMSD) between structures, with an option to include hydrogen atoms or not. Our algorithm is implemented with Python and can efficiently obtain minimized RMSD taking into account of the symmetry of molecular systems , contributing to the fields of drug design, molecular docking, and data analysis of molecular dynamics simulation.

## INTRODUCTION

The Root Mean Square Deviation (RMSD) is an important metric for quantifying the dissimilarity of the three-dimensional structures of two conformers. This metric finds broad application in fields including chemical informatics, crystal structure analysis, drug design, and molecular dynamics, where it plays a pivotal role in comparing diverse molecular geometries and deciphering conformational variations.

Given two molecules or clusters in three-dimensional structures – a reference structure denoted as α and a target structure denoted as β – the practical execution of the RMSD algorithm typically proceeds through the following steps:

1. Ordering of Atomic Correspondences. This is a fundamental prerequisite for a meaningful RMSD calculation, necessitating a one-to-one matching of atomic arrangements between the α and β.
2. Centering on a Common Geometric Origin. Both the α and β molecular frameworks undergo a translational shift as a whole, aligning their geometric centroids precisely on the coordinate system's origin.
3. Rotation for Optimal Alignment and RMSD Computation. A search ensues for an optimal rotation matrix which, when applied to the coordinate vectors of structure β, minimizes the RMSD between the coordinates of α and β. The resulting minimal RMSD value thus obtained signifies the true measure of spatial discrepancy between the two molecules.

Step 2 of the process is the most straightforward, involving merely calculating the centroids and subtracting these from each atom's coordinates. Step 3, widely recognized in the fields of chemistry, biology, and crystallography as the Kabsch algorithm [1], traces its origins in the literature to Kabsch's work in 1976 [1-2]. The quaternion method has also been demonstrated to be equivalent to the Kabsch algorithm when applied in Step 3 [2-3]. While Step 1 has often been overlooked, or considered trivial. However, the common way used by computers to process a molecule in 3D structure is based on the sequential arrangement of the constituent atoms with spatial coordinates in the input files. For a single case, reassigning labels to match the order of atoms between a pair of small molecules is a manageable task. As the number of atoms per molecule increases, or when dealing with a large number of molecules for comparison, it is no longer a trivial matter. Moreover, the challenge is exacerbated when molecules contain a significant number and variety of topologically equivalent atoms, rendering Step 1 a highly complex endeavor.

Canonization algorithms address this by systematically ordering the constituent atoms according to their inherent connectivity and properties, which is fundamental for generating a unique molecular representation[4]. The unique representation facilitates searching and managing molecules in a database, assigning effective registration numbers to newly-found compounds, and constructing hash codes for molecules to store in a database. Over the past several decades, many canonization algorithms have been proposed [4-17], each with its own rules and objectives such as searching chemical compound files and substructures [18-19], doing pharmacophore alignment [20], and deriving unique identifiers for molecules[3, 5, 6], etc. Examples include the Weininger algorithm[5] implemented in the commercial software Daylight, the classical Morgan algorithm[6], and the International Chemical Identifier (InChI) endorsed by IUPAC[7-8]. Most of these algorithms work with the one-dimensional (1D) using string or two-dimensional (2D) using connectivity graph data of a molecule, and are primarily designed to handle constitutional and topological information but may not account for three-dimensional (3D) stereochemistry explicitly. So they are not suitable for generating one-to-one mapping between two 3D structures of different conformers.

The stereochemical tag on a chiral atom is in fact determined according to the established Cahn-Ingold-Prelog (CIP) rules[21-22] for comparing grades of the different branches. This helps to further distinguish similar atoms into different parts. In 2015, Schneider, Sayle, and Landrum (SSL) proposed a novel canonization algorithm implementing a stable index assignment strategy and an efficient partition-based graph relaxation method [10]. This algorithm is robust and efficient, marking a significant advance towards the goal. However, the SSL algorithm is not compatible with the CIP nomenclature[21-22], and nor is the SSL algorithm capable of accurately comparing structural differences between two 3D conformations of a molecule with different atomic orders. The reason is that for topologically symmetric molecules with the same chiral tags exhibiting distinct spatial arrangements of their symmetric components, variations necessarily exist in the atom-to-atom mapping relations. When those topologically symmetric atoms or groups maintain a symmetric spatial disposition yet differ in the sequential mapping of atoms to one another, this leads to divergent RMSD outcomes.

Coutsias, E.A., et al., developed a program named frmsd[23] using C and Fortran languages to deal with the issue of symmetric atoms posing a challenge in minimized RMSD computations, which employs quaternions. Upon provision of a list of symmetric atoms by the user, this program iterates over the all possible permutations, ultimately outcomes the minimal RMSD. Rocco Meli and colleagues introduced spyrmsd, a Python-based program [24], that leverages the VF2 algorithm [25] to ascertain isomorphisms in molecular graphs, thereby identifying all groupings of symmetric atoms. It then computes RMSDs for all possible permutations, retaining the configuration with the lowest RMSD. To our current knowledge, spyrmsd stands as the most comprehensive tool for automatically calculating the accurate lowest RMSD given a pair of molecular graph structures and atomic coordinates. The authors of spyrmsd emphasized that the graph isomorphism problem is a non-polynomial (NP) problem, hence the symmetry-corrected RMSD calculations are only suited for molecules of modest to intermediate sizes. [24]

In this report, a novel canonization algorithm is proposed that retains merits of the SSL algorithm, such as the stable index assignment strategy and the graph relaxation method, and improves the original algorithm through redesigning the initial assignment rules to fit CIP rules as far as possible. Furthermore, it recruits a new branching tiebreaking step to obtain minimized RMSD between two 3D structures of the same molecule or conformers, as well as the real one-to-one atom mapping. The canonization along with the branching tiebreaking algorithms help to reduce the time-scaling needed for minimized RMSD.

The algorithm has been implemented using Python and RDKit module [26], and is available at GitHub (https://github.com/JerryJohnsonLee/CanonizedRMSD)[27].

## METHOD AND IMPLEMENTATION

**Overview**. Graph relaxation methods, such as the Weisfeiler-Lehman Procedure[28], are widely used for determining whether two graphs are isomorphic [29]. Chemical structures can be regarded as undirected weighted graphs with atomic properties, including atomic number, isotopic atomic weight, charge, and the 3D coordinates of the atom encoded in the nodes. [28] Therefore, we employed a graph relaxation method for canonizing molecules and took advantage of the abundant information encoded in the nodes.

Our canonization algorithm proceeds as follow: (1) the atoms are firstly assigned to initial indices according to their local properties, and then (2) the refinement step uses graph relaxation method to update the index of each node iteratively on the basis of direct neighbors. If the molecule is not highly symmetrical (which is almost always the case), these two steps will assign the same index to atoms with the same chemical environment and assign different indices to those in different environments (See Section A and B in the Supporting Information (SI) for a proof). After (3) stereochemical assignment to chiral or isomeric atoms accompanied with further refinement, (4) the final tiebreaking step is performed to designate unique index to each atom. Scheme 1 depicts the flowchart of the whole procedure, and each of the basic steps will be elaborated in the following sections.

**Scheme 1.** Flowchart describing the basic steps of our algorithm.

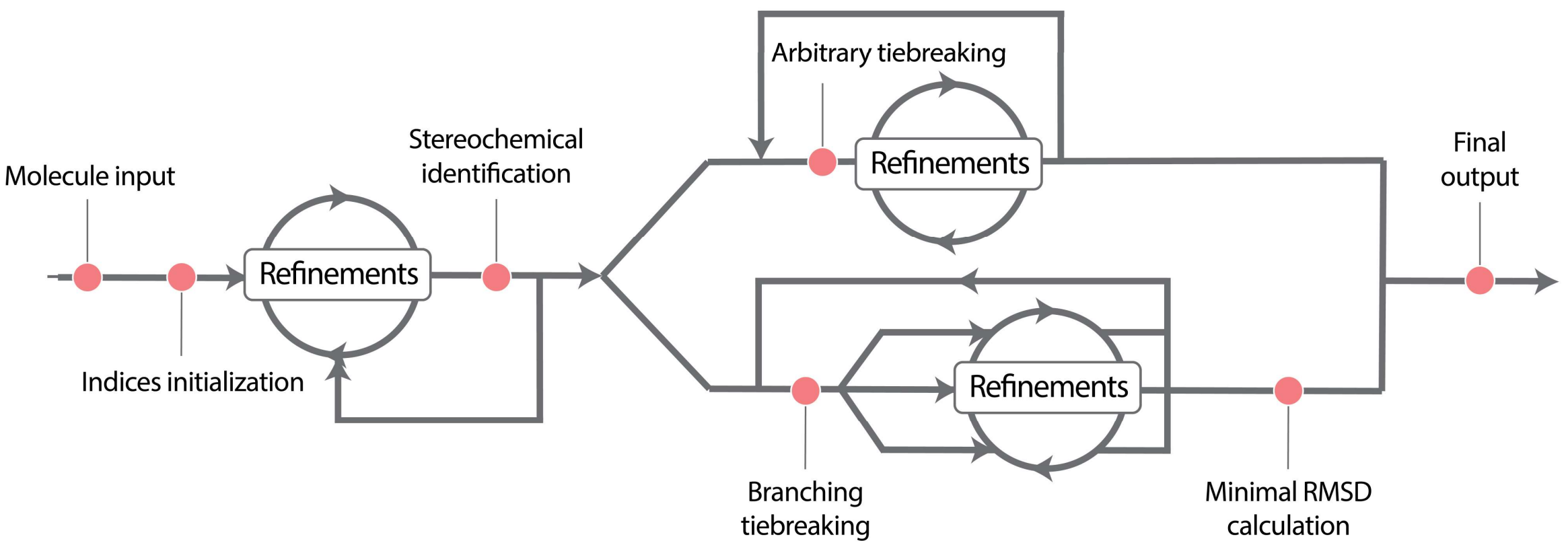

**Initial Assignment.** The initial assignment step aims to attribute atoms into as many partitions as possible in terms of their local properties as is done in the SSL algorithm [10], while each “partition” contains atoms sharing the same CI and local properties. If requiring canonization on the non-hydrogen structure, then all hydrogen atoms were first removed before initial assignment. Otherwise, hydrogen atoms were retained and were treated in the same way as non-hydrogen atoms. To obtain CIP-compatible CIs, a new initial assignment step is adopted constituting the following three rules:

1. Atoms with higher atomic numbers acquire higher indices.
2. Atoms with heavier adjacent atoms acquire higher indices.
3. Atoms with higher coordination numbers (CNs) acquire higher indices.

Rules 1 and 2 explicitly agree with two of the CIP rules, while CN in rule 3 is defined as $\mathrm{CN} = \mathrm{degree} + N_{\mathrm{double\ bond}} + 2 \times N_{\mathrm{triple\ bond}} + 0.5 \times N_{\mathrm{aromatic\ bond}}$, also for CIP-compatibility. The degree stands for the total number of non-hydrogen atoms connected to the central atom. When double bonds or triple bonds are encountered, CIP rules regard the atoms at the remote end as duplicated or triplicated, respectively, and treat the new atoms as imaginary ones [21]. The expression of CN takes into account the contributions from these imaginary atoms with the additive terms in order to ensure compatibility. If two atoms have the same atomic number, adjacent atoms and CN, their precedence cannot be determined immediately, and their order will be decided by the neighboring atoms in the subsequent refinement step.

As the rules are executed sequentially, more atomic properties are considered, which generates more partitions. Notably, instead of assigning consecutive indices to different partitions, just the right amount of the CI space is reserved for each partition, so that the CIs could be stable as far as possible and vary only within the reserved scope in the following steps. Specifically, if there are $n_{\mathrm{X}}$ atoms in the partition assigned with index X, the indices for atoms in the next partition would be $X + n_X$. Owing to the index stability, atoms sharing no indices with other atoms could be instantly marked as finalized with its index fixed till the end of the entire process.

(a)

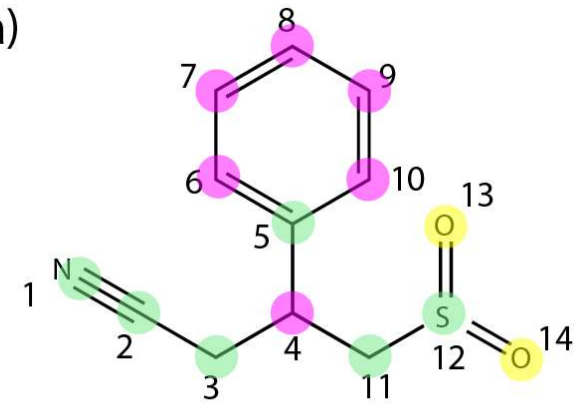

molecule **1**

(b)

| original index | 1 | 2 | 3 | 4 | 5 | 6 | 7 | 8 | 9 | 10 | 11 | 12 | 13 | 14 |
|---|---|---|---|---|---|---|---|---|---|---|---|---|---|---|
| rule 1 | 11 | 1 | 1 | 1 | 1 | 1 | 1 | 1 | 1 | 1 | 1 | 14 | 12 | 12 |
| rule 2 | 11 | 9 | 1 | 1 | 1 | 1 | 1 | 1 | 1 | 1 | 10 | 14 | 12 | 12 |
| rule 3 | 11 | 9 | 1 | 2 | 8 | 2 | 2 | 2 | 2 | 2 | 10 | 14 | 12 | 12 |

**Figure 1.** (a) Structure of model molecule 1 (4-(dioxo-$\lambda^5$-sulfanyl)-3-phenylbutanenitrile). Atoms are colored according to their states after the initial assignment step ends. Atoms with the same color (other than green) are of the same partition and share the same canonical index under the applied rule(s). Green color denotes finalized atoms that share no indices with other atoms. The original index of each atom is marked with black number. (b) Change history of CIs as the three rules are applied sequentially. Each column stands for one atom. Color scheme used for grids is the same as that for atoms in (a).

Figure 1 illustrates the change history of CIs of model molecule **1** during the initial assignment step. All hydrogen atoms are excluded from the atomic graph for simplicity, leaving 14 atoms to be canonized. Following rule 1, only 4 types of atoms can be distinguished: one sulfur atom (original index 12), two oxygen atoms (original indices 13,14), one nitrogen atom (original index 1) and ten carbon atoms (original indices 2-11). While the carbon atoms are assigned the lowest index, the principle of index stability dictates that the nitrogen atom which has the second lowest atomic number obtains a CI of 11 (=10+1) to leave enough index space for carbon, two oxygen atoms share a CI of 12, and the sulfur atom is indexed 14. Rule 2 enables further discrimination of atoms 2 and 11 because they are connected with atoms other than carbon, respectively. Atom 11 has higher CI (10) than atom 2 (9), featuring its heaviest adjacent atom of sulfur, which has larger atomic mass weight than the nitrogen atom next to atom 2. CN is taken into consideration when rule 3 is executed. In all the remaining atoms with CI of 1, atom 3 has the lowest CN and atom 5 has the highest, whereas the other carbon atoms remain indistinguishable. Therefore atom 3 attains the lowest CI in the CI space of the partition (1), atom 5 acquires the highest (8), and the CIs of the other carbon atoms are updated to 2, which completes the initial assignment step for the molecule.

**Refinement.** The refinement step further differentiates atoms and divides the partitions into smaller ones. This step adopts a graph relaxation method and runs iteratively. Every iteration deals with one partition containing atoms currently sharing the same CI. This partition-based strategy decreases the computational complexity by focusing on a finite number of atoms in each iteration. When there are multiple partitions, these partitions are proceeded sequentially following the 'longer partition first, higher canonical index second' principle. In the selected partition, a descending index list of immediate neighbors (DILIN) is generated for each atom, and these atoms are then sorted accordingly. The 'Towers of Hanoi' algorithm [9] is recruited here for sorting because of its high efficiency in processing data with many duplicate keys. Atoms with DILINs of larger indices will be assigned to larger CIs, which divides the selected partition into smaller ones. The CI space is still retained for each partition for the sake of index stability. Then a new iteration is started with the top-ranked partition on the basis of renewed CIs.

A reservoir set is maintained during the refinement step to keep track of all the atoms that need to be considered and is used to determine when refinement is accomplished. The reservoir set is initially filled with all the atoms whose CIs are not finalized. After each iteration, the atoms in the investigated partition are removed from the set. But if there are index updates in the iteration, the unfinalized direct neighbors of any of the updated atoms and all the atoms sharing the same partition with these neighbors will be added to the reservoir. The iterations continue until the reservoir set is cleared.

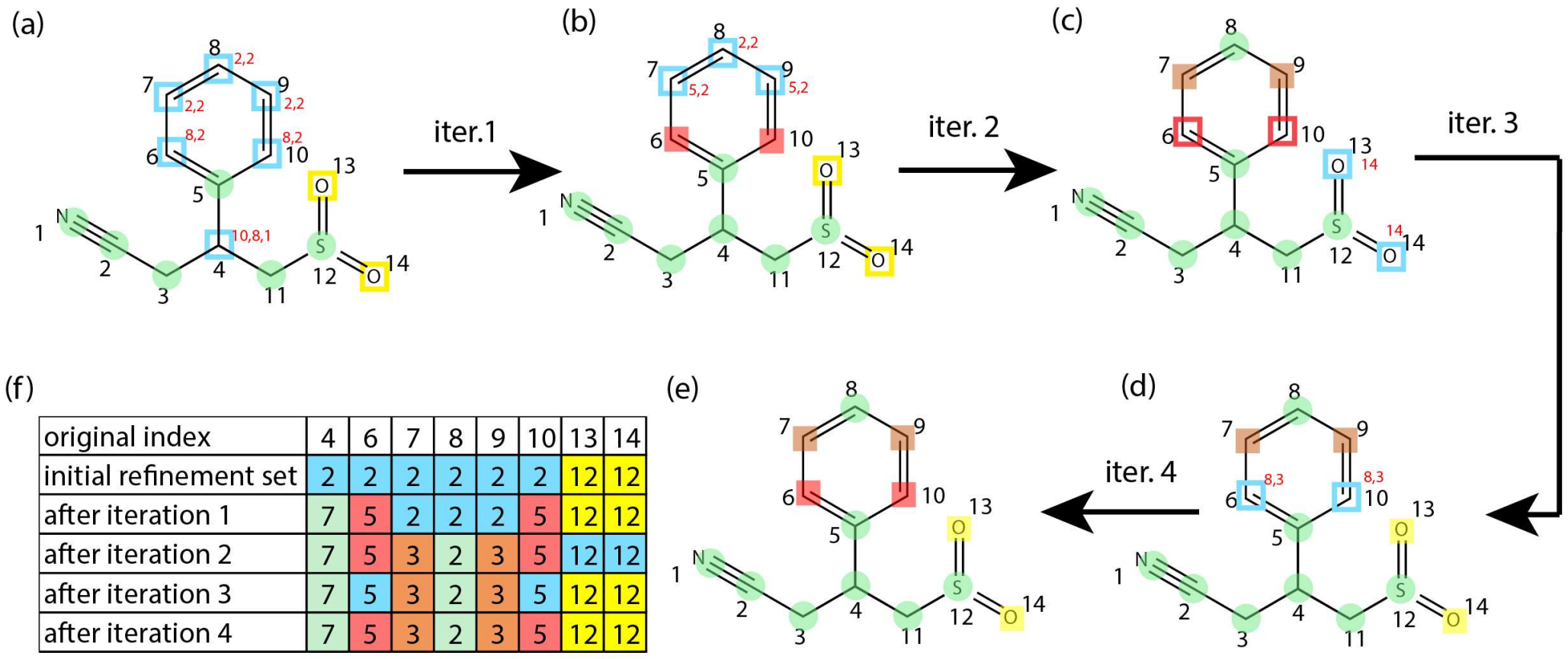

| original index | 4 | 6 | 7 | 8 | 9 | 10 | 13 | 14 |
|---|---|---|---|---|---|---|---|---|
| initial refinement set | 2 | 2 | 2 | 2 | 2 | 2 | 12 | 12 |
| after iteration 1 | 7 | 5 | 2 | 2 | 2 | 5 | 12 | 12 |
| after iteration 2 | 7 | 5 | 3 | 2 | 3 | 5 | 12 | 12 |
| after iteration 3 | 7 | 5 | 3 | 2 | 3 | 5 | 12 | 12 |
| after iteration 4 | 7 | 5 | 3 | 2 | 3 | 5 | 12 | 12 |

**Figure 2.** (a-e) Diagram of the refinement step for model molecule 1. Colors of red, orange, yellow, and blue are used to differentiate atoms in different partitions and shape symbols are used to represent the states of atoms before the next iteration starts. Green filled circles denote finalized atoms, whereas filled or hollow squares denote unfinalized ones. Hollow squares are used for atoms in the reservoir set and the blue hollow ones are particularly dedicated to the atoms in the investigated partition. The descending index lists of immediate neighbors (DILIN) of the investigated atoms are given in red numbers. (f) Change history of CIs in the refinement step. Grids use the same color scheme as that for atoms in (a-e).

Figure 2 illustrates the entire procedure of refinement step on model molecule **1**. The reservoir set initially contains 8 atoms, which belong to 2 partitions. The first iteration deals with the longer partition in which the atoms share a CI of 2 (Figure 2a and 2f). The DILIN of each atom is derived subsequently. Atom 4 connects to atoms 3, 5 and 11, whose current CIs are 1, 8 and 10 (Figure 2b), respectively, generating a DILIN of 10,8,1. Atom 6 and 10 share a DILIN of 8,2 and atom 7, 8 and 9 have a common DILIN of 2,2. Accordingly, the selected partition splits into three sub-partitions. Atom 4 is finalized with the largest number in the CI space (7) due to its highest DILIN priority, the CIs for atoms 6 and 10 are updated to 5 due to their second priority, and the CIs of atom 7, 8 and 9 remain to be 2 (Figure 2b). The reservoir set is renewed by firstly removing all the atoms in the partition (atoms 4, 6, 7, 8, 9 and 10) and then bringing back atoms 7, 8 and 9 which are in a sub-partition adjacent to atoms 6 and 10, whose CIs are updated in the current iteration. Hence, the reservoir set now contains atoms 7, 8, 9, 13 and 14. In the next iteration, the investigated atoms still have a CI of 2, involving atoms 7-9 (Figure 2b). Atom 8 is differentiated from atoms 7 and 9 by its unique DILIN (2, 2) and is finalized with a CI of 2, while the CIs of atoms 7 and 9 are updated to 3 in the same iteration (Figure 2c and 2f). Due to the update, atoms 6 and 10 next to atoms 8 and 9 are brought back to the reservoir set after the removal of atoms 7, 8 and 9. The third iteration focuses on the two oxygen atoms as all unfinalized partitions are of the same length but theirs have the highest CI (Figure 2f). However, the oxygen atoms share the same DILIN and cannot be further distinguished. Removal of the oxygen atoms from the reservoir set is the only change of this iteration (Figure 2d and 2f). The last iteration behaves similarly as it changes nothing but clears the reservoir set (Figure 2e and 2f). Overall, the refinement step splits the original 2 large partitions into 5 smaller ones.

**Stereochemical Assignment.** When stereochemistry matters in canonization, all chiral and isomeric atoms are identified and assigned with stereochemical tags in this step. As the former two steps ensure that the atoms of different CIs are chemically non-identical, a chiral atom can be easily identified as a tetrahedral atom connecting three or four non-hydrogen atoms with diverse CIs. Furthermore, our algorithm ensures that the CIs of the neighboring atoms are ordered according to the CIP rules (see Section A in the SI for further explanation), and the stereochemical configuration can thus be easily identified. In practice, the configuration is derived after calculating the sign of the scalar triple product spanned by the vectors connecting the central atom and/or the neighboring atoms as illustrated in Figure 3a and 3b. Supposing that the four neighboring atoms marked with a, b, c and d are arranged in descending order by their CIs, and C is the central atom, the sign of the scalar triple product of vectors $\overrightarrow{ab}$, $\overrightarrow{ac}$ and $\overrightarrow{Ca}$ ($\overrightarrow{ab} \times \overrightarrow{ac} \cdot \overrightarrow{Ca}$) determines the stereochemistry. A positive sign corresponds to **S** configuration of the central atom, whereas a negative sign corresponds to **R** configuration. Furthermore, our algorithm also allows a straightforward determination of stereochemical configuration of double bond from CIs, which is defined by the sign of the inner product of $\overrightarrow{ab}$ and $\overrightarrow{cd}$ ($\overrightarrow{ab} \cdot \overrightarrow{cd}$): positive for **Z** configuration and negative for **E**. The neighboring atoms are arranged *a priori* to ensure that atom a is prior to b and c is prior to d in terms of CIs (Figure 3c and 3d).

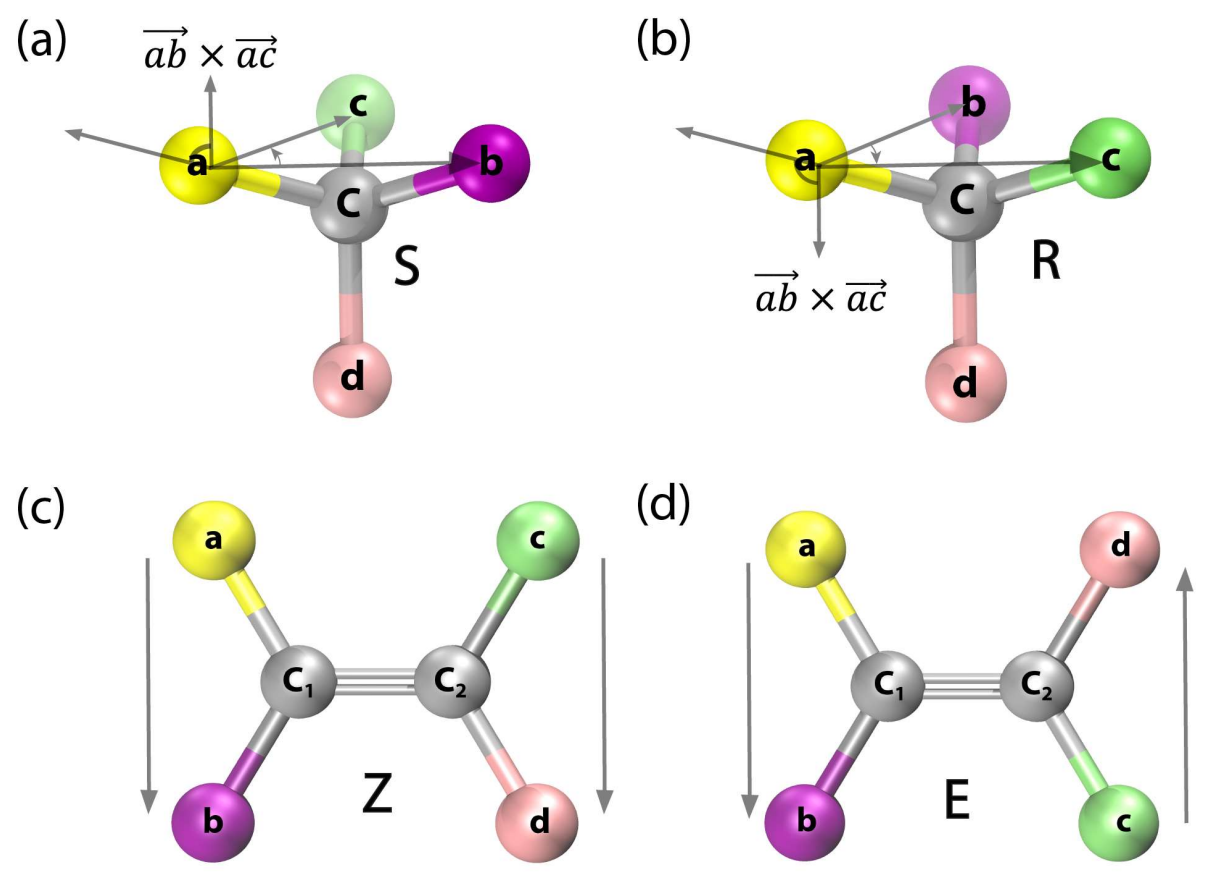

**Figure 3.** After the iterative refinement, the atoms with bigger CIs are precedential according to the CIP rules. The stereochemistry tag of an atom or a bond can therefore be assigned according

to the coordinates and CIs of its adjacent atoms. In (a) and (b), the neighboring atoms marked with a, b, c and d are in descending order of CIs. In (c) and (d), the neighboring atom a is prior to b, and c is prior to d in terms of CIs.

The identification of the stereochemical configuration of chiral atoms or double bonds enables further discrimination of nearby atoms. Specifically, atoms of a *cis*- double bond or having **R** configuration are updated with higher CIs, and the indices for atoms of a *trans*- double bond or having **S** configuration have lower CIs. When a stereochemical assignment step finishes, a subsequent refinement step is performed (see Scheme 1) to update the indices affected by stereochemistry identification. While the stereochemical assignment focuses on chiral or isomeric atoms, the refinement step iteratively differentiates direct and indirect neighbors of these atoms and confers different CIs to them. Note that stereochemical assignment-refinement combination needs to run iteratively to ensure that all the pseudo-chiral atoms (whose neighboring atoms are discriminated by their chirality) are analyzed, and the r/s chiral tags instead of R/S tags are assigned to these atoms [30].

**Tiebreaking.** The ultimate goal of canonization is to assign unique indices to every atom. After the above three steps, however, there may still be atoms with equivalent CIs and the tiebreaking step is employed to warrant uniqueness. If two molecules are compared only in terms of their graph topology, atoms with equivalent CIs are intrinsically the same (see Section B in the SI for a proof). Under this circumstance, the indices can be arbitrarily assigned to atoms within the available CI space (Scheme 1). However, when the 3D coordinates of atoms are taken into consideration, the atoms with equivalent CIs are no longer the same. Here we recruit the branching tiebreaking step to exhaust all the possible canonizations to ensure completeness, similar to the backtrack tree proposed before[1]. Specifically, a breadth-first branching tree of the to-be-tiebroken molecule is generated, and the structure of the tree is decided by the 'longer partition first, higher index second' principle. For a selected partition that contains $n_{\mathrm{X}}$ atoms sharing a CI of X for tie-breaking, the algorithm enumerates every possibility of assigning a unique lowest index for one of the atoms which made its index finalized, and assigns all the remaining atoms in the partition with +1 indices. The resulting tree will have $\mathrm{n_X}$ separated child branches. A refinement step is then performed on each of the branch to search for potentially discriminable atoms due to the splitting of partition and confers different CIs to them if possible. The refinement step could avoid branching as much as possible and decrease the total number of canonizations. When comparing the root-mean-square deviation (RMSD) between two molecules, a RMSD minimization step is performed here based on all atoms with unique CIs for $\mathrm{n_X}$ branches to pinpoint the minimum-RMSD assignment. Following that branch, the branching-refinement-RMSD-calculation combination continues if there are other atoms that do not have unique CIs and require further tie-breaking. It should be noted that the branching tiebreaking step may be very time-consuming, especially for highly symmetrical molecules that have massive number of possible canonizations. Therefore this step is performed only when necessary, such as calculating the minimal RMSD between two conformations of a molecule. However, for molecules containing atoms with the same local chemical environment, the SSL algorithm assigns basically random CIs to these atoms [10], which risks overestimating the value of RMSD between the conformations and misrecognizing two similar conformations as very different ones. Such an issue is solved with the current algorithm (see Section C in the SI for more discussion).

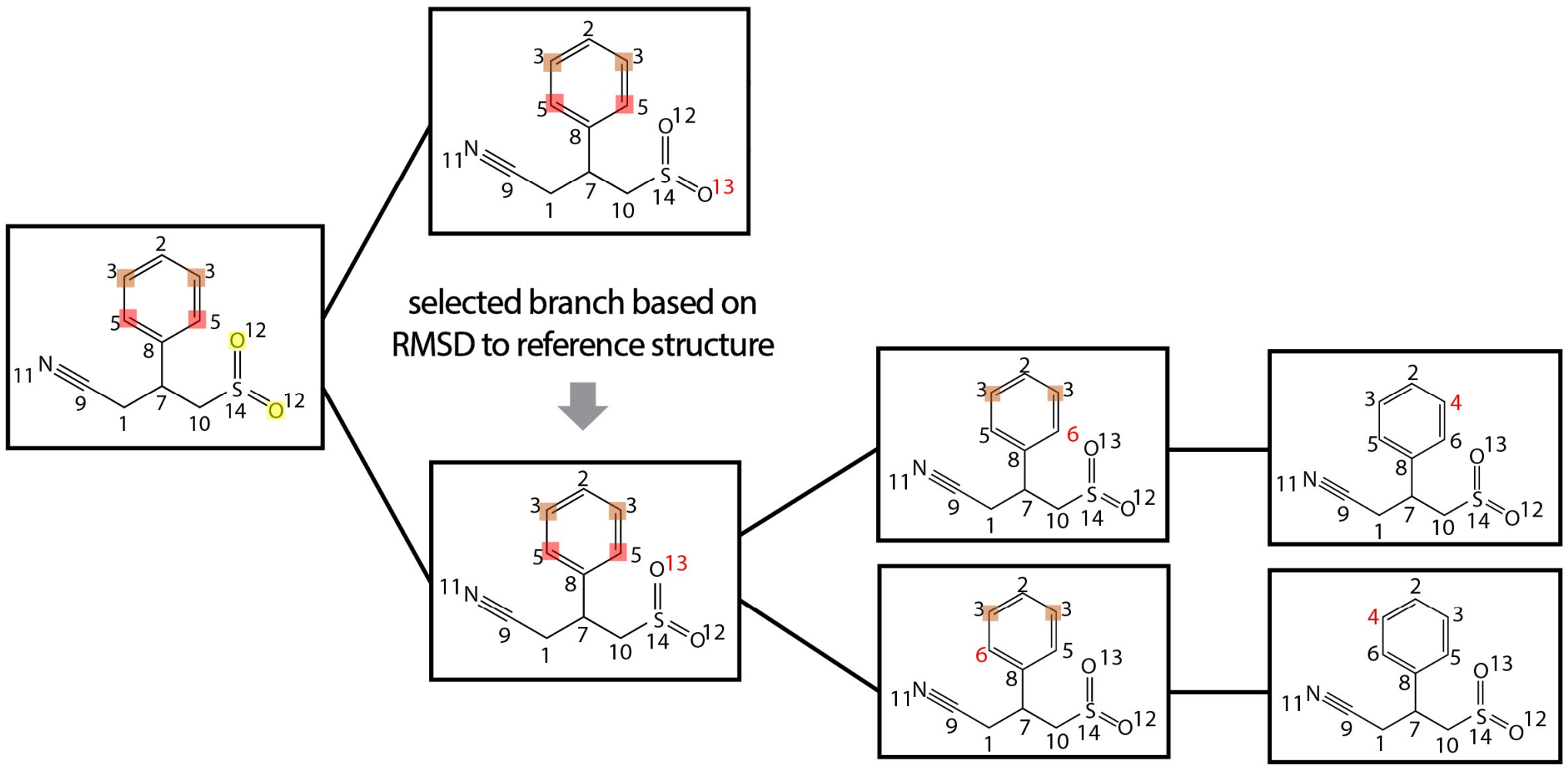

**Figure 4.** The breadth-first branching tree generated by tiebreaking step. Red numbers indicate the updated CIs after branching or refinement, and each leaf node represents a possible canonization. Atoms in the same partition are denoted by the same symbols of the same color.

Figure 4 illustrates the branching tiebreaking procedure for model molecule **1**. Since all unbroken partitions are of the same length, the tiebreaking step starts with the partition of the largest CI, which contains two oxygen atoms. Two permutations, 12,13 and 13,12, are assigned to these two atoms, respectively, leading to two branches in the tree, though the subsequent refinement step does nothing. Based on the availability of the 3D geometry of a reference structure, one of these two branches will generate lower RMSD and only that branch will be further expanded. In the next cycle, the two atoms indexed 5 are broken and are assigned to 5 and 6, respectively, again resulting two canonizations. The latest branching also affects two atoms indexed 3, as these atoms are distinguished by the following refinement step, which avoids generating new branches in the tree. Comparing RMSDs with the reference structure once more on these two canonizations will decide the global minimal RMSD as well as the optimal mapping between atoms in two conformations.

**Duplicate Checking.** Duplicate checking is an important application of canonization, aiming to determine whether two structures are of the same molecule. It is also a prerequisite for structural comparison before calculating the minimized RMSD between two structures. As the checking is done only for graph connectivity, it firstly canonizes the two original molecules with arbitrary tiebreaking after the previous steps, and then the atoms in each molecule are rearranged according to the obtained canonical indices. The next step is an atom-by-atom comparison, checking whether the pair of atoms sharing the same index has exactly the same atomic properties and the same neighbor atoms with the same bond orders. Any tiny difference in above features indicates structures of non-identical molecules, otherwise, the structures are of the same molecule. Moreover, a string may be condensed from canonical indices, atomic properties, and bond orders to act as a unique identification code for the molecule when necessary.

**Minimized RMSD with Linked Hydrogens.** The above rules are handled with the assumption that H atoms are ignored. Since the indices of the H atom can be placed after all the heavy atoms, the corresponding index partitions can be assigned according to the indices of the connected heavy atom, especially based on the indices minimized by the RMSD of the heavy atom skeleton. Finally, for the locally symmetric degenerate H atom groups, the minimum RMSD is calculated by implementing full permutation (temporarily fixing the indices of other groups) in each group to determine the order of all the H atoms in the group, while avoiding the coupling of different permutation with other groups. Take **Figure 5.** as an example:

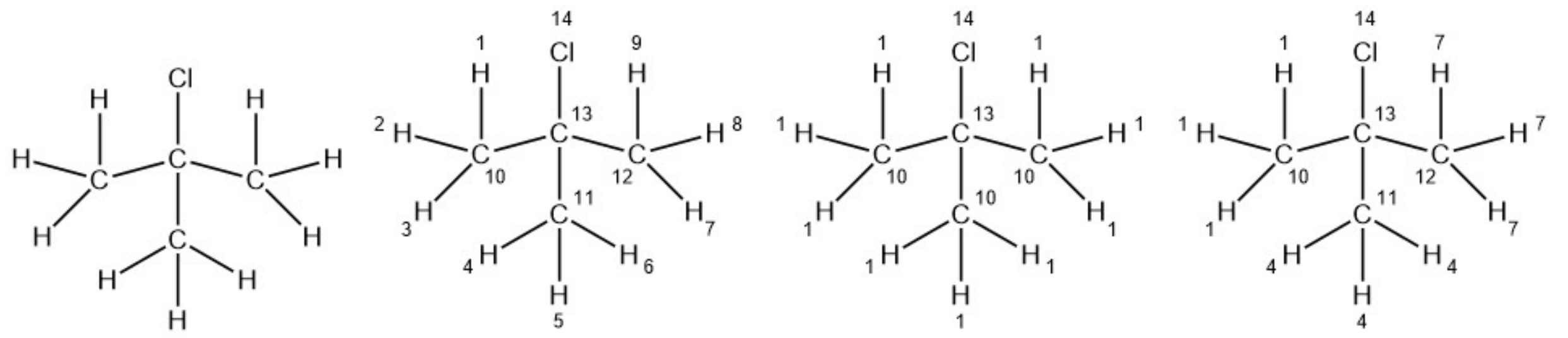

(a) (b) (c) (d)

**Figure 5**: Demonstration diagrams in the permutation and tiebreaking process. (a) The molecular diagram before canonization. (b) The canonized reference molecule. Its C atoms with No. 10-12 are the topologically equivalent atoms with arbitrarily assigned indices. The corresponding linked H atoms are grouped into different partitions of the corresponding order, and the H atoms in each group continue to be assigned arbitrary indices to obtain the final order. (c) The canonized target molecule. The topologically equivalent atoms are grouped separately and assigned the minimum value in the number segment. (d) The canonization step of the target molecule after calculating the minimum RMSD corresponding to the reference molecule only with non-hydrogen atom skeleton, the corresponding indices of C atoms with No. 10-12 is obtained, and then the linked H atoms are grouped according to the indices of its linked non-hydrogen atom.

For complex molecules, the algorithm pruned the branches in time when searching at each layer of the tree, selected the node that minimized RMSD at this time based on all atoms with finalized indices, and then only continued to search the subtree of this node. Since the search of the whole tree is avoided, the computation amount is greatly reduced. For example, for the molecules used in the example in **Figure 5**, the number of RMSD calculations required by the conventional traversal method reaches $3! \times (3! \times 3! \times 3!) = 1296$ times, and the number of calculations of the current algorithm is only $3! + (3! + 3! + 3!) = 24$ times.

## RESULTS AND DISCUSSION

**Evaluation of Structural Difference.** The new branching tiebreaking step employed in our algorithm enables an accurate evaluation of differences between two structures. Five testing cases were prepared and each contains two structures (Figure 6). The molecules range from small organic molecules with several atoms to large biomolecule containing more than 500 atoms. For cases a and b, the two structures used for comparison come from the same molecule and share identical conformation, but one of them is randomly translated and rotated and the original atomic indices are scrambled for testing purpose (Figure 6a-b). Our algorithm correctly identifies that the structures are of the same molecule in both cases through duplicate checking and finds that every pair of structures is the same with a minimal RMSD of zero (Table 1). In case c-e, two structures with slightly different conformations are used and our algorithm distinguishes the minor differences between the structures. Notably, very different conclusions may be obtained when the arbitrary tiebreaking step of the SSL algorithm is used.

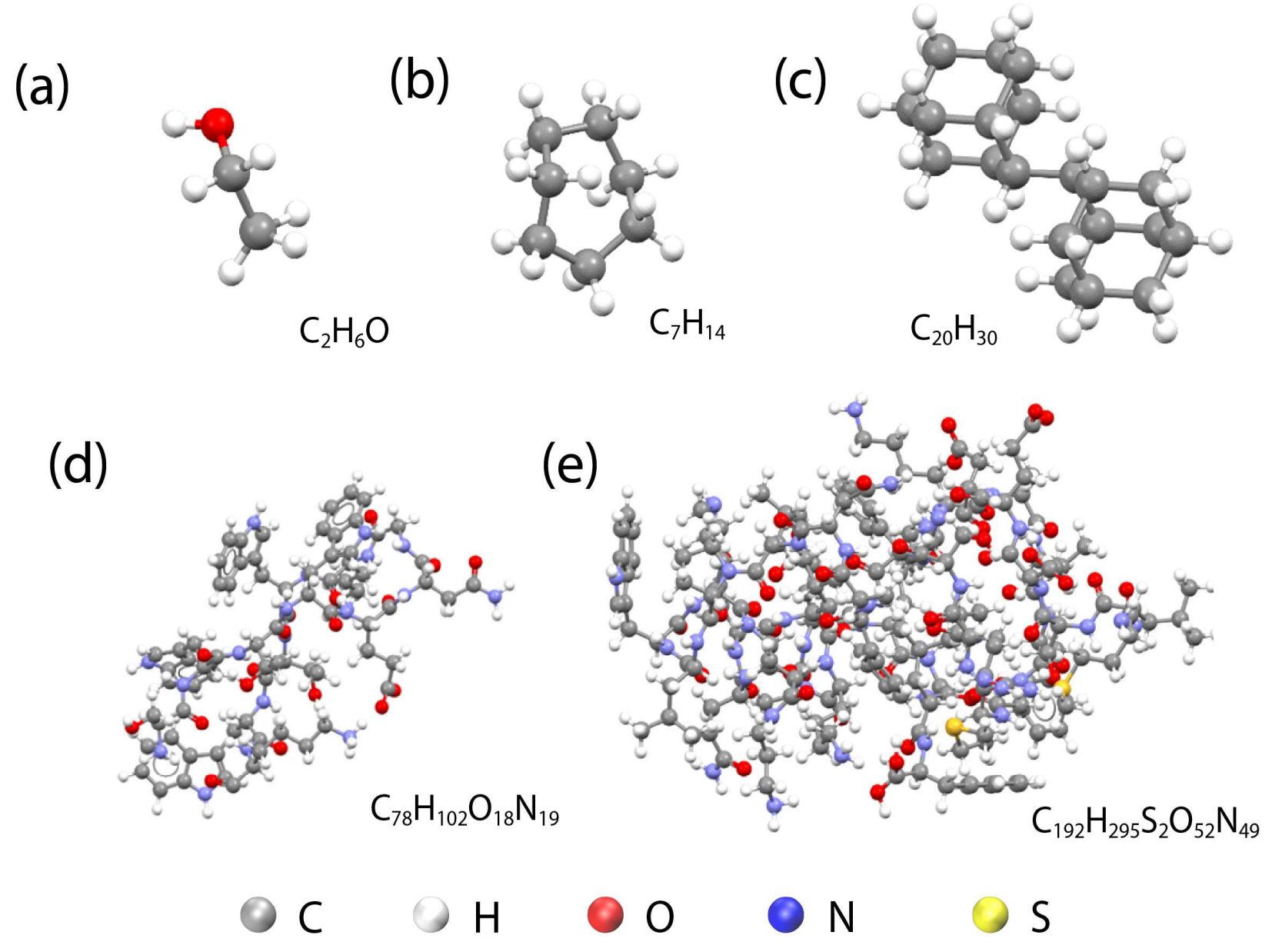

**Figure 6.** Structures of the molecules used as testing cases. The structure files in SDF and PDB formats can be found in the SI.

**Table 1.** Calculation of the minimal RMSD for five testing cases.

| Molecule index | Number of atoms in molecule | Minimal RMSD (Å)[a] | Running time for this work (s) [b] | Running time for **spyrmsd** [24] (s)[c] |
|---|---|---|---|---|
| a | 9 | 0.000 | 0.35 | 0.77 |
| b | 21 | 0.000 | 0.48 | 1.24 |
| c | 50 | 0.086 | 0.74 | 38.13 |
| d | 217 | 1.614 | 9.85 | \ |
| e | 590 | 1.698 | 389.44 | \ |

Each case was run 10 times and the runtime is given as an average in the table. All H atoms are included to calculate minimal RMSD. The computation was done on one core of a workstation with Intel Xeon Platinum 8269C CPU which has a clock rate of 2.5 GHz.

[a] Minimum of the RMSDs obtained from all possible canonizations after optimized rotation and translation of the structures.

[b] The atom sequence has been changed randomly before calculate RMSD in each time.

[c] “\” means the running time of the code is too long (more than 10 minutes) or encountering an error so no results can be obtained.

Table 1 summarizes the running time required for each calculation, while the results are compared with the most recent RMSD calculation package, **spyrmsd** [24] . Our algorithm shows better time-scaling respect to the molecule size and can deal with much bigger molecule within acceptable time. Although **frmsd** [23] developed with C and Fortran runs more quickly for calculating the optimal alignment to minimize RMSD, it requires an extra file from the user to define the indices for atoms with the same atom type, while this information is automatically inferred in our algorithm and spyrmsd. Our algorithm works much better on molecules with lower symmetry, since it scales down the possible branches that a high symmetry structure needed. Fortunately, molecules with extremely high symmetry are quite rare in real applications. Finally, we would like to emphasize that our algorithm still has reasonable running time for large biomolecules such as molecule d and e, which takes about 10 seconds and 6 minutes, respectively. Our program includes an option to calculate minimized RMSD with or without considering hydrogen atoms, a detail that is often overlooked by default in many other programs. Considering symmetrically equivalent hydrogen atoms, which are prevalent in organic molecules, can significantly increases the computational complexity. Our algorithm handles these permutations efficiently. By restricting the permutation of equivalent hydrogen atoms to within groups connected to the same atom, we drastically reduce the combinatorial explosion, thereby making routine calculations for large molecules possible and affordable.

We further tested the algorithm on n-alkanes, water molecular clusters, and other molecules with various chiral structures and high symmetries. In the following test, the coordinate order of all the atoms in the reference molecule $\alpha$ (including hydrogen atoms) was randomly shuffled and saved as the target β molecule, and then the lowest RMSD between α and β were calculated. So the standard answer should be 0 definitely. The results on the minimum RMSD calculations and the running times for both our work (CanonizedRMSD) and spyrmsd were provided in Supporting Tables 1-3 for the n-alkanes, water clusters and highly symmetrical molecules respectively. Due to the arrangement of the equivalent atoms in the graph, the RMSD results calculated by spyrmsd are all 0, showing excellent stability. However, when there are a few more equivalent atoms in the graph, the computational efficiency decreases sharply. It takes 0.87s and 2.29s to compute n-butane and n-octane respectively, and the time for calculating n-$C_{12}H_{26}$, n-$C_{14}H_{30}$ and n-$C_{16}H_{34}$ has risen to 23.87s, 157.32s and 594.55s. The rapid increase in computation time rendered testing of larger molecules impractical. Our algorithm demonstrates a significantly lower computational scaling, taking only goes from 0.40 s for n-butane to 31.50s for n-$C_{80}H_{162}$, showing a much lower computational scale. Meanwhile, the RMSDs calculated by our algorithm were also all zeros, demonstrating the correctness of our results.

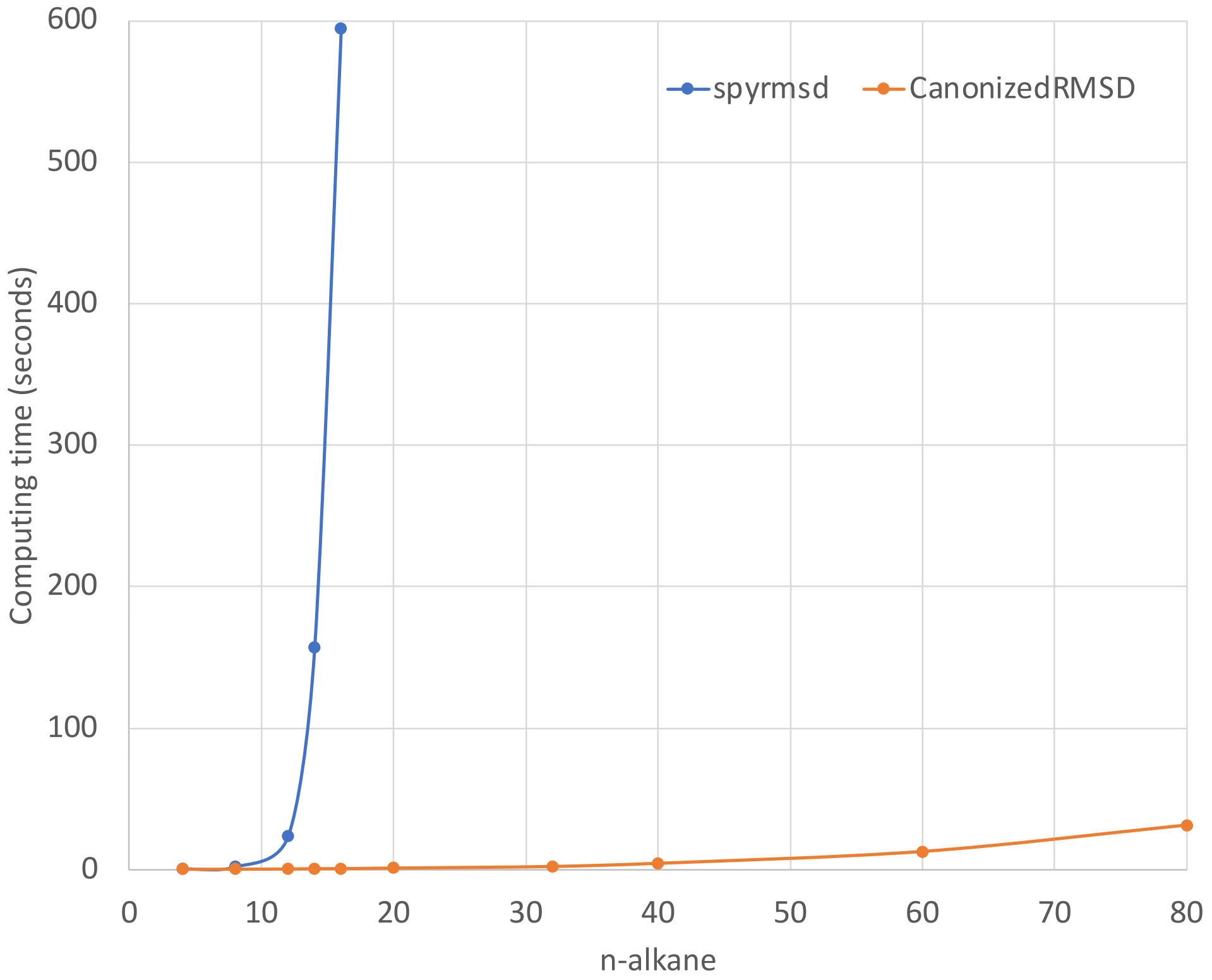

**Figure 7.** The calculation time of RMSD for n-alkanes vs the number of carbon atoms.

Similarly, both programs can calculate RMSD for clusters of water molecules. Because there is no bonding relationship between each water molecule, the algorithm can only carry out full permutation processing for the molecular slices without bonding relationship. This is quite an extreme case of algorithmic time. Spyrmsd processed the cluster of eight water molecules in 2095.74 seconds. Thanks to the group pruning algorithm of H atoms, our algorithm (CanonizedRMSD) shows a very low computation scale, and can process 100 water molecular clusters within 40 minutes, with a total of 300 atoms, and get a perfect matching situation with RMSD of 0.

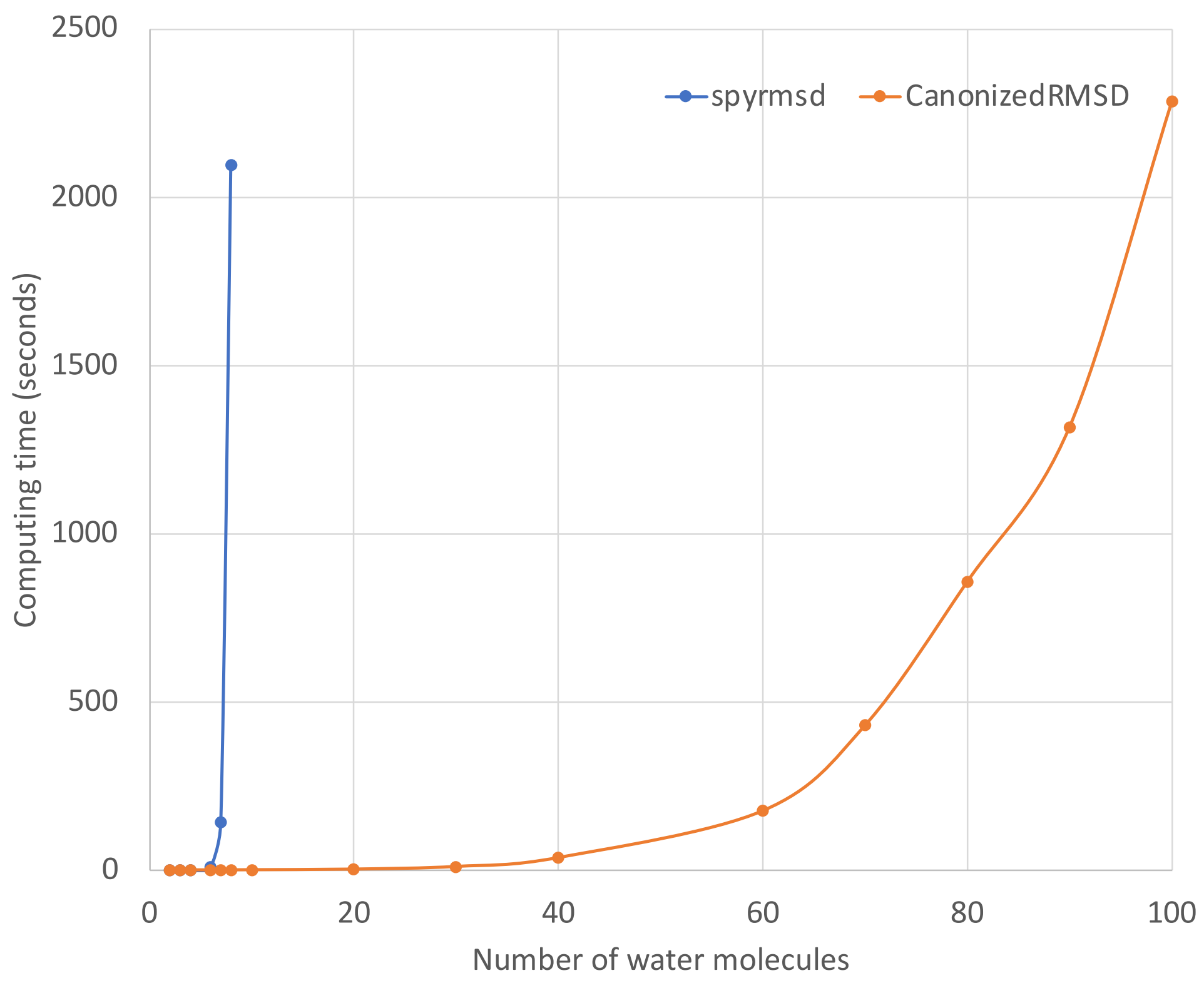

**Figure 8.** The calculation time of RMSD for water clusters vs the number of water molecules.

**Kekule structure.** For aromatic ring compounds commonly used in organic molecules, rdkit recognizes aromatic bonds when reading molecular structures built with different Kekule structures of the same compound, which will be automatically considered to be the same molecular topology in this python program.

**Stereochemical assignment.** Stereochemical assignment of chiral atoms could be easily obtained by our program and stereochemistry tags can be output when required. The basis for chirality determination, which inherently involves comparing the grade of adjacent atoms or groups, enhances our capability to canonize molecular structures through the identification of chiral atoms. We employed a set of 300 molecular structures proposed by Hanson et. al. [31], encompassing various known possibilities of chirality as well as all examples from the IUPAC BlueBook V3 [32]. The results are summarized in Supporting Table 4, including calculated RMSD values, running time by our method, running time by spyrmsd, and the RMSD calculation result using an arbitrary canonization with the SSL method [10]. Our algorithm consistently achieved a 100% success rate with RMSD=0 across these structures. While Spyrmsd managed to compute correct outcomes within reasonable durations for most cases, it exceeded a 15-minute threshold and was manually terminated for more than 30 larger molecules or those with a higher number of equivalent atoms.

Although our algorithm does not yet perfectly adjudicate every CIP rule, the approximation of CIP-based size determinations significantly reduces symmetry degeneracy in complex molecules and lessens the computational load required for achieving the lowest RMSD. Additionally, inconsistencies in the application and interpretation of CIP rules across different molecular simulation software were observed during our research. Six molecules (cases g-l) were used here for testing (Table 2). All the stereochemistry tags obtained by our algorithm are in good consistency with those derived from the CIP rules. We also tested three widely used chemical informatics toolkits, including GaussView [33], RDKit [26] and ChemOffice on the benzene-based molecules. Though GaussView and ChemOffice provide correct answers to four cases, RDKit fails in two of them, cases g and i (Table 2), indicating that the CIP rules are not closely followed by the toolkits.

**Table 2.** Comparison of stereochemistry tags generated by different programs.

| Case | Structure | Chiral atom index | Stereochemistry tag | | | | |
|---|---|---|---|---|---|---|---|
| | | | CIP | Our algorithm | GaussView 5[1] | RDKit[2] | Chem3D 20[3] |
| g | 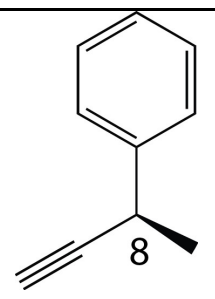 | 8 | S | S | S | R | S |
| h | 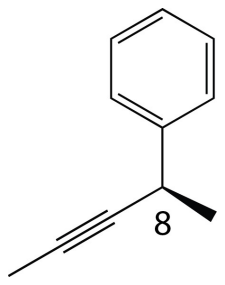 | 8 | **R** | R | R | R | R |
| i | 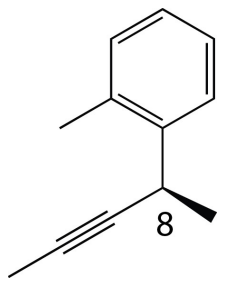 | 8 | **S** | S | S | R | S |
| j | 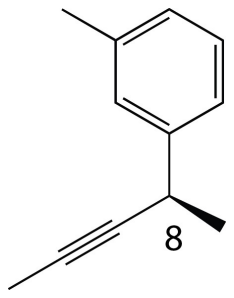 | 8 | **R** | R | R | R | R |
| k | 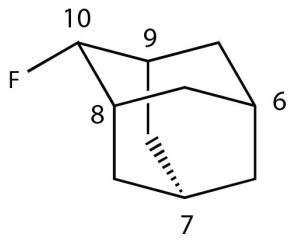 | 6 | \ | \ | S | \ | r |
| | | 7 | \ | \ | R | \ | r |
| | | 8 | \ | \ | r [4] | \ | S |
| | | 9 | \ | \ | s [4] | \ | R |
| | | 10 | \ | \ | R | \ | \ |
| l | 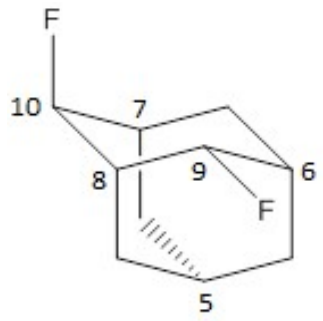 | 5 | **r** | r | S | R | S |
| | | 6 | **S** | S | S | S | S |
| | | 7 | **R** | R | R | R | R |
| | | 8 | **s** | s | S | S | R |
| | | 9 | **S** | S | S | S | S |
| | | 10 | **S** | S | S | S | S |

Tags in red color are wrong. Tags in blue color do not differentiate pseudo stereochemistry. The 3D coordinate files of these molecules can be found in the GitHub repository[27] and in the SI.

[1] GaussView Version 5.0.8

[2] RDKit Python conda Version 2023.9.6 using rdkit.Chem.AssignStereochemistryFrom3D

[3] Chem3D Ultra Version 20.1.1.125

[4] These pseudo chiral tags cannot be shown with GaussView Version 6.1.1.

Cases k and l belong to another kind of molecule attributed to substituted adamantane. Case k has fluorine substitution at only one position, which retains the mirror symmetry of the molecule. Although there is essentially no chirality in any atom of the molecule, only our algorithm and RDKit correctly report this fact, whereas GaussView and Chem3D fail by assigning stereochemistry tags for five of the carbon atoms (Table 2). Case l was created by adding two substituent groups at the asymmetrical positions, which breaks the original molecular symmetry and confers stereochemistry to six of the carbon atoms. Though the stereochemical assignment of our algorithm is well in line with the CIP rules, all the three chemical informatics toolkits fail for pseudo-chiral atoms 5 and 8. All in all, the test cases demonstrate the power of our algorithm in deriving CIP-compatible CIs that are useful for stereochemical assignment, whereas the tested chemical informatics toolkits sometimes make assignment against the CIP rules.

**CONCLUSION**

Current molecular canonization methodologies often rely on 1D string representations or 2D connectivity graphs that inherently do not take advantage of the explicit 3D stereochemical information. We proposed a novel multifunctional molecular canonization algorithm here on the basis of the SSL algorithm, taking into account of 3D information. After the redesign of the initial assignment rules, the algorithm greatly simplifies the identification of CIP stereochemistry tags for chiral atoms and double bonds, that enhances its canonization ability to distinguish atoms into meticulous partitions. And the introduction of branching tiebreaking step facilitates the accurate evaluation of the minimal RMSD between two 3D structures. All those techniques when combined greatly reduces the possible permutation to deal with molecules with a considerable number of symmetric atoms and the corresponding calculation time scaling.

The algorithm has already been implemented in Python, with the RDKit module to import molecules and attain graph connections. The program is available at GitHub, and can be easily used for acquiring the canonization mapping from the original indices to the CIs, for generating a unique identification code for each molecule, or be integrated into other programs in multiple fields.

It should be noted that our algorithm may fail for molecules with extremely high symmetry, such as clusters of water molecules without explicit bonding information. However, considering that this kind of molecules is quite rare, our algorithm is still robust in most cases. For example, we computed the RMSD including hydrogen atoms for all 19 highly symmetrical molecules mentioned in Figure 8 of reference [16], successfully replicating a 100% match in each instance. Our next work would be to extend the applicability of the algorithm to those extremely highly symmetrical molecules and periodic systems.

## ASSOCIATED CONTENT

**Supporting Information**. The following files are available free of charge.

The SI comes with compatibility description of the canonical indices (CIs) with CIP rules, proof of uniqueness of canonized CIs for equivalent atoms and discussion about minimizing the root-mean-square deviation with equivalent atoms (SI.pdf)

ZIP package of all structure files in format of SDF/PDB/MOL for molecules a-l (SI_structures.zip)

Xlsx worksheet for Supporting Tables 1-4 (SI_tables.xlsx)

AUTHOR INFORMATION

**Corresponding Author**

Jianming Wu – Department of Chemistry, Fudan University, Shanghai, 200438, China; Email: jianmingwu@fudan.edu.cn

**Present Addresses**

†If an author's address is different than the one given in the affiliation line, this information may be included here.

**Author Contributions**

The manuscript was written through contributions of all authors. Jianming Wu directs this project. Jie Li and Qian Chen write the codes and perform tests. All authors have given approval to the final version of the manuscript.

**Funding Sources**

This project is supported by the National Science Foundation (22233002, 21373053) and the Science Challenge Project (TZ2018004).

# Supporting information for

## Canonized then Minimized RMSD for Three-Dimensional Structures

Jie Li, Qian Chen, Jingwei Weng, Jianming Wu*, Xin Xu

(Department of Chemistry, Fudan University, Shanghai, 200438, China)

**Section A. Compatibility of the canonical indices (CIs) with CIP rules.**

Our algorithm provides CIs whose orders are in compliance with the CIP rules and the explanation is provided below.

The CIP rules for ordering the neighboring groups of a central atom are given as follows [1]:

**Sub-rule 0:** Near end or side precedes far.

**Sub-rule 1:** Higher atomic number precedes lower.

**Sub-rule 2:** *Cis* precedes *trans*.

**Sub-rule 3:** **R** precedes **S** (for pseudo-asymmetry only).

**Sub-rule 4:** Higher mass-number precedes lower (among isotopes only).

Among the sub-rules, four of them are explicitly stipulated by our algorithm. Sub-rules 1 and 4 are equivalent to rules 1 and 2 of our initial assignment step, respectively, and sub-rules 2 and 3 aiming at chiral atoms or double bonds are warranted by our stereochemical assignment step.

Sub-rule 0 is the exception as it does not explicitly appear in any of the rules we define, and instead it is implicitly guaranteed by the refinement step. Supposing that we have a central tetrahedral atom, two of whose direct neighboring atoms are of the same CI after the initial assignment step finishes, the refinement step would then explore each of the groups led by these two atoms, respectively, to see whether the two groups can be discriminated from each other or not. As the iterations of refinement step proceed, atoms whose DILINs are of higher priority will be assigned to larger CIs. The decisive role of DILIN means that the priority of the leading atoms is determined by the nearer atoms rather than the farther ones. In the case of stereochemical assignment, if two groups of the leading atoms do have discrepancies, the nearest discrepancy identified in the initial assignment step actually determines. As the priority of atoms at the discrepancy is compatible with CIP rules, the refinement step will propagate this priority through DILIN to the leading atoms, and sub-rule 0 is thusly guaranteed.

All in all, by implementing the rules defined in the initial assignment step and the refinement step, the CIs generated by our algorithm is in compliance with the CIP rules and could be directly used for ordering the neighboring groups of a central atom or a double bond during stereochemical assignment.

### Section B. Proof of uniqueness of canonized CIs for equivalent atoms

If two atoms are in the same chemical environment, these two atoms should be identical, both in local properties and in overall bond topologies, i.e. all the side chains starting from these two molecules need to be exactly the same [2]. In practice, all the local atomic properties in addition to atomic number, including isotopic atomic weight and atomic charge are included during the initial index assignment. That ensures all atoms having the same indices share the same local properties. For atoms in a chain, all the side chains starting from these atoms must also be identical. That is because if there is a deviation between two chain branches, the atoms where the deviation occurs must be different in local properties, or they must be connecting to other atoms by different bond types. If the first situation is correct, the two atoms where the deviation first occur are naturally given the different indices according to the explanation before. If the second situation is true, we then need to consider whether these atoms have the same degree. If they have the same degree, that means the number of branches these atoms connect to are the same, but the bond orders are different. Then by definition, the coordination numbers of these atoms are different, and they are given different initial indices at first place. Conversely, if the degrees of the atoms where deviation occurs are not the same, these atoms will be put into different partitions during the refinement, because the lengths of their INIL are different. The consequence is that the atoms where deviation occurs will always obtain different indices after refinement. Then since the deviation in atomic indices will propagate all the way back to the origin of the branches during the iterative refinement, all the atoms in different branches will be given different indices, and that means only atoms in the same chemical environment will obtain the same index after refinement.

## Section C. Minimizing the root-mean-square deviation

Molecular canonization could contribute to quantifying the structural difference between two conformations of a molecule which is routinely encountered in molecular docking and rational drug design studies [3-4]. Given two sets of vectors representing the coordinates of atoms in two conformations, respectively, the classical Kabsch's algorithm [5] or the more efficient quaternion-based characteristic polynomial (QCP) algorithm [6] can be used for generating a best rigid-body movement to minimize the root-mean-square deviation (RMSD) between two conformations. Smaller RMSD indicates more similar conformations. However, a prerequisite for applying the Kabsch's or QCP algorithm is that the optimal atom-to-atom mapping is known in advance as explicitly represented by the vectors through the order of atomic coordinates [7], otherwise, the alignment algorithm could not warrant the minimal value of RMSD. At the first sight, the SSL canonization algorithm could solve the problem by mapping atoms through their CIs in two molecules. However, for molecules containing atoms with the same local chemical environment, the SSL algorithm assigns basically random CIs to these atoms [8], which risks overestimating the value of RMSD between the conformations and misrecognizing two similar conformations as very different ones. Figure S1 illustrates two essentially identical neopentane structures that could in principle align with each other perfectly and acquire an RMSD of zero as long as the correct mapping relationship (a to a', b to c', c to b' and d to d') is determined. However, since the four terminal carbon atoms a-d are in the same chemical environment, the SSL algorithm makes no use of structural information of these atoms and assigns random CIs to them. Therefore the obtained CIs may lead to wrong atom-to-atom mapping and these two identical conformations may be misrecognized as different ones. By introducing branching tiebreaking based on local structural alignments, our algorithm is able to tackle this issue.

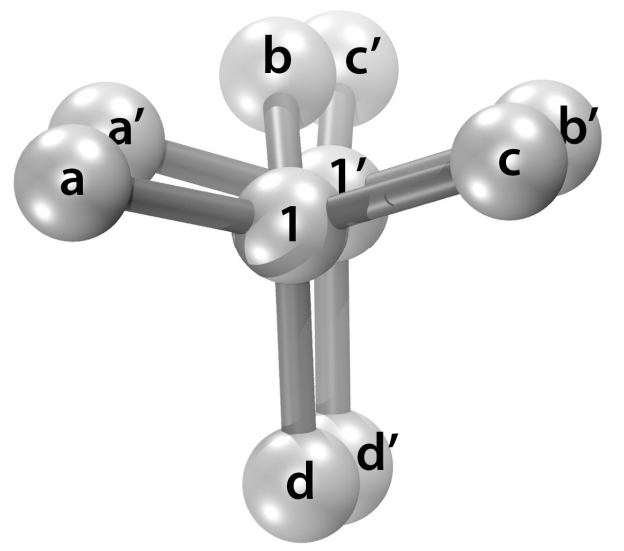

**Figure S1.** Example of two neopentane structures that non-optimal atom-to-atom mapping (b→b' and c→c') generated for atoms with the same chemical environment leads to overestimation of the value of RMSD and misrecognition of the identical structures as different ones. Hydrogen atoms are neglected for the sake of simplicity in this example.